\documentclass[11pt]{article}
\usepackage{amssymb}
\usepackage{amsmath}
\usepackage{amscd}
\usepackage{latexsym}
\usepackage{bbm}

\def\a{\alpha}
\def\b{\beta}
\def\ga{\gamma}
\def\la{\lambda}
\def\ga{\gamma}
\def\de{\delta}
\def\eps{\epsilon}
\def\ve{\varepsilon}

\def\Si{\Sigma}

\def\vp{\varphi}
\def\th{\theta}
\def\O{\Omega}

\newcommand{\R}{\mathbb R}

\newcommand{\Gcal}{{\cal G}}
\newcommand{\Acal}{{\cal A}}

\newcommand{\Mcal}{{\cal M}}
\newcommand{\Fcal}{{\cal F}}
\newcommand{\Ncal}{{\cal N}}
\newcommand{\Tcal}{{\cal T}}

\newcommand{\N}{{\mathbb N}}

\newcommand{\gfrak}{{\mathfrak g}}

\newcommand{\ah}{{\hat{\smash{a}}}}
\newcommand{\bh}{{\hat{\smash{b}}}}
\newcommand{\hd}{{\hat{\textrm{d}}}}

\def\im{\textrm{i}}

\def\diff{\textrm{d}}
\def\pa{\mbox{$\partial$}}
\def\sfrac#1#2{{\textstyle\frac{#1}{#2}}}
\def\+{\dagger}
\def\={\ =\ }

\def\and{\quad\textrm{and}\quad}
\def\with{\quad\textrm{with}\quad}
\def\for{\quad\textrm{for}\quad}

\def\Id{\mathrm{Id}}

\begin{document}

\begin{titlepage}
\setcounter{page}{0}
\begin{flushright}
ITP-UH-12/15
\end{flushright}

\hspace{2.0cm}

\begin{center}

{\Large\bf Green-Schwarz superstring as subsector of Yang-Mills theory }

\vspace{12mm}

{\large  Alexander D. Popov}\\[8mm]

\noindent {\em
Institut f\"ur Theoretische Physik \\
Leibniz Universit\"at Hannover \\
Appelstra\ss e 2, 30167 Hannover, Germany }\\
{Email: popov@itp.uni-hannover.de
}\\[6mm]

\vspace{10mm}

\begin{abstract}
\noindent We consider Yang-Mills theory with $N=2$ super translation group in ten auxiliary dimensions as the structure group.
The gauge theory is defined on a direct product manifold $\Sigma_2\times H^2$, where $\Sigma_2$ is a two-dimensional Lorentzian
manifold and $H^2$ is the open disc in $\R^2$ with the boundary $S^1=\partial H^2$. We show that in the adiabatic limit, when
the metric on  $H^2$ is scaled down, the Yang-Mills action supplemented by a Wess-Zumino-type term becomes the Green-Schwarz
superstring action.
\end{abstract}

\end{center}
\end{titlepage}

\noindent {\bf 1. Introduction. } Superstring theory has a long history~\cite{1}-\cite{3} and pretends on description of all
four forces in Nature.  On the other hand, Yang-Mills theory (plus matter fields) in four dimensions describes three main
forces except the gravitational force. The aim of this short paper is to show that the Green-Schwarz superstring theory (of
type I, IIA and IIB)~\cite{12} can be  obtained as a subsector of pure Yang-Mills theory with a Lie supergroup  $G$ as the
structure group. In fact, we introduce a Yang-Mills model on a direct product manifold $\Si\times\widetilde\Si$ which in the
low-energy limit flows to the Green-Schwarz superstring model on $\Si$, with $N{=}2$ super translation group $G$ in $d{=}10$
as a target space. Here $\Si$ and $\widetilde\Si$ are two-dimensional Lorentzian and Riemannian manifolds, respectively. Our
construction is based on the results of papers~\cite{1a}-\cite{4a}, where it was shown that $\Ncal{=}4$ super-Yang-Mills
theory on $\Si\times\widetilde\Si$ flows to a sigma-model on $\Si$ while $\widetilde\Si$ shrinks to a point. The target space
of the above sigma-model is the space of vacua that arise in the compactification on $\widetilde\Si$. For pure Yang-Mills
theory on $\Si\times\widetilde\Si$ this is the moduli space of flat connections on $\widetilde\Si$~\cite{1a}-\cite{4a}. In our
paper we reverse the logic of ~\cite{1a}-\cite{4a} and associate a Yang-Mills model on $\Si\times\widetilde\Si$ to the known
superstring model on $\Si$. We consider the simplest case of the super translation group $G$ as the structure group but other
supergroups can be considered as well, e.g. the group $G=\,$PSU$(2,2|4)$ related with the coset space AdS$_5\times S^5$.

\bigskip

\noindent {\bf 2. Lie supergroup $G$.}  We consider Yang-Mills theory on a direct product manifold $M^4=\Si_2\times H^2$,
where $\Si_2$ is a two-dimensional  Lorentzian manifold (flat case is included) with local coordinates $x^a, a,b,...=1,2$, and
a metric tensor $g^{}_{\Si_2}=(g_{ab})$, $H^2$ is the open disc with coordinates $x^i$, $i,j,...= 3,4$ and the metric tensor
$g^{}_{H^2}=(g_{ij})$. Then $(x^\mu )=(x^a, x^i)$ are local coordinates on $M^4$ with metric tensor $(g_{\mu\nu})= (g_{ab},
g_{ij} ), \mu , \nu = 1,...,4$.

As the structure group of Yang-Mills theory, we consider  the coset $G{=}$SUSY$(N{=}2)/SO(9,1)$ (cf. \cite{4}) which is the
subgroup of $N{=2}$ super Poincare group in ten auxiliary dimensions generated by translations and $N{=}2$ supersymmetry
transformations. Its generators $(\xi_{\a}, \xi_{Ap})$ obey the Lie superalgebra $\gfrak =\,$Lie$\,G$,
\begin{equation}\label{1}
\{\xi_{Ap}, \xi_{Bq}\} =(\ga^\a C)_{AB}\de_{pq}\xi_\a\ ,\quad [\xi_{\a}, \xi_{Ap}]=0\ ,\quad [\xi_{\a}, \xi_{\b}]=0\ ,
\end{equation}
where $\ga^\a$ are the $\ga$-matrices, $C$ is the charge conjugation matrix, $\a = 0,...,9,\ A=1,...,32$ and $p,q=1,2$ label
the number of supersymmetries.  Coordinates on $G$ are $X^\a$ and two spinors $\th^{Ap}$ of the
Majorana-Weyl type. On the superalgebra $\gfrak=\,$Lie$\,G$ we introduce the metric $\langle\cdot \rangle$ with components
\begin{equation}\label{2}
\langle\xi_{\a}\, \xi_{\b}\rangle=\eta_{\a\b}\ , \quad \langle\xi_{\a}\, \xi_{Ap}\rangle=0\and \langle\xi_{Ap}\, \xi_{Bq}\rangle=0\ ,
\end{equation}
where $(\eta_{\a\b})=\,$diag$(-1,1,...,1)$ is the Lorentzian metric on $\R^{9,1}$ and the last equality in (\ref{2}) is
standard in superstring theory.

\bigskip

\noindent {\bf 3. Yang-Mills action.} We consider the gauge potential $\Acal =\Acal_{\mu}\diff x^\mu$ with values in $\gfrak$
and the $\gfrak$-valued  gauge field
\begin{equation}\label{3}
 \Fcal =\sfrac12\Fcal_{\mu\nu}\diff x^\mu \wedge \diff x^\nu\with \Fcal_{\mu\nu} =\partial_\mu\Acal_\nu - \partial_\nu\Acal_\mu
 + [\Acal_\mu , \Acal_\nu]\ ,
\end{equation}
where $[\ ,\ ]$ is the commutator or anti-commutator for two $\xi_{Ap}$-generators. On  $M^4=\Si_2\times H^2$ we have the
obvious splitting
\begin{equation}\label{4}
 \diff s^2 = g_{\mu\nu}\diff x^\mu \diff x^\nu = g_{ab}\diff x^a \diff x^b + g_{ij}\diff x^i \diff x^j\ ,
\end{equation}
\begin{equation}\label{5}
\Acal =\Acal_{\mu}\diff x^\mu= \Acal_{a}\diff x^a+\Acal_{i}\diff x^i\ ,
\end{equation}
\begin{equation}\label{6}
 \Fcal =\sfrac12\Fcal_{\mu\nu}\diff x^\mu \wedge \diff x^\nu =\sfrac12\Fcal_{ab}\diff x^a \wedge \diff x^b + \Fcal_{ai}\diff x^a \wedge \diff x^i
+\sfrac12\Fcal_{ij}\diff x^i \wedge \diff x^j\ .
 \end{equation}

By using the adiabatic approach in the form presented in \cite{5, 6}, we deform the metric (\ref{4}) and introduce
\begin{equation}\label{7}
 \diff s^2_\ve = g_{\mu\nu}^\ve\,\diff x^\mu \diff x^\nu = g_{ab}\diff x^a \diff x^b + \ve^2g_{ij}\diff x^i \diff x^j\ ,
\end{equation}
where $\ve\in [0,1]$ is a real parameter. Then $\det (g_{\mu\nu}^\ve )=\ve^4\det (g_{ab}) \det (g_{ij})$ and
\begin{equation}\label{8}
 \Fcal^{ab}_\ve = g_\ve^{ac}g_\ve^{bd}\Fcal_{cd}= \Fcal^{ab}\ ,\quad \Fcal^{ai}_\ve = g_\ve^{ac}g_\ve^{ij}\Fcal_{cj}=
 \ve^{-2}\Fcal^{ai}\and
 \Fcal^{ij}_\ve = g_\ve^{ik}g_\ve^{jl}\Fcal_{kl}=\ve^{-4}\Fcal^{ij}\ ,
\end{equation}
where indices in $\Fcal^{\mu\nu}$ are raised by the non-deformed metric tensor $g^{\mu\nu}$. It is assumed that  $\Fcal^{\mu\nu}$
smoothly depend on $\ve$ with well-defined limit for $\ve\to 0$.

For the deformed metric  (\ref{7}) the Yang-Mills action functional is
\begin{equation}\label{9}
S_\ve=\frac{1}{2\pi}\,\int_{M^4} \diff^4x\,\sqrt{|\det g^{}_{\Si_2}|}\,\sqrt{\det g^{}_{H_2}}\,\left\{\ve^2\langle\Fcal_{ab}\,
\Fcal^{ab}\rangle + 2\langle\Fcal_{ai}\, \Fcal^{ai}\rangle + \ve^{-2}\langle\Fcal_{ij}\, \Fcal^{ij}\rangle\right\}\ ,
\end{equation}
where $\pi$ is the ``area" of the disc $H^2$ of radius $R=1$.

\bigskip

\noindent {\bf Remark.} On the disc $H^2$ of radius $R=1$ one can consider both the flat metric $g_{ij}=\de_{ij}$ (then
${\mbox{Vol}(H^2)}=\pi$) and the metric
\begin{equation}\label{10}
g_{ij}=\frac{4}{(1-r^2)^2}\,\de_{ij}\with r^2=\de_{ij}x^ix^j\ .
\end{equation}
However, we will see later that in all integrals over $H^2$ the metric $g_{ij}$ enters in the combination $\sqrt{\det
g^{}_{H_2}}\,g^{ij}\xi_i\xi_j=\de^{ij}\xi_i\xi_j=1$, where $(\xi_i)=(\sin\vp, -\cos\vp )$ is the unit vector on $S^1=\pa H^2$.
Hence all calculations for the metric  (\ref{10}) are equivalent to the calculations  for $g_{ij}=\de_{ij}$. That is why we
will consider the flat metric on $H^2$ as in many mathematical papers considering Yang-Mills theory on the balls $B^n$ with
$n\ge 2$.

\bigskip

\noindent {\bf 4. Adiabatic limit.} The term $\ve^{-2}\langle\Fcal_{ij}\, \Fcal^{ij}\rangle$ in the Yang-Mills action
(\ref{9}) diverges when
 $\ve\to 0$. To avoid this we impose the flatness condition
\begin{equation}\label{11}
 \Fcal_{ij}=0
\end{equation}
on the components of the field tensor along $H^2$ for $\ve =0$. However, for $\ve>0$ the condition (\ref{11}) is not needed
and one can consider $ \Fcal_{ij} (\ve>0)\ne 0$, only $\Fcal_{ij}(\ve =0)= 0$.

In the adiabatic limit $\ve\to 0$, the Yang-Mills action (\ref{9}) becomes
\begin{equation}\label{12}
S_0=\frac{1}{\pi}\int_{M^4} \diff^4x\,\sqrt{|\det g^{}_{\Si_2}|}\,\langle\Fcal_{ai}\, \Fcal^{ai}\rangle
\end{equation}
with the equations of motion
\begin{equation}\label{13}
D_i\Fcal^{ib}:=\frac{1}{\sqrt{|\det g^{}_{\Si_2}|}}\,\partial_i\left(\sqrt{|\det
g^{}_{\Si_2}|}\,\de^{ij}g^{ab}\Fcal_{aj}\right) + [\Acal_i , \Fcal^{ib}]=0\ ,
 \end{equation}
\begin{equation}\label{14}
D_a\Fcal^{aj}:=\frac{1}{\sqrt{|\det g^{}_{\Si_2}|}}\,\partial_a\left(\sqrt{|\det
g^{}_{\Si_2}|}\,\de^{ij}g^{ab}\Fcal_{ib}\right) + [\Acal_a , \Fcal^{aj}]=0\ .
 \end{equation}
Note that the metric $g^{}_{\Si_2}$ on $\Si_2$ is not fixed and the Euler-Lagrange equations for $g^{}_{\Si_2}$ yield the
constraint equations
\begin{equation}\label{15}
T_{ab}^0 = \de^{ij} \langle\Fcal_{ai}\, \Fcal_{bj}\rangle -\sfrac12
 g_{ab}\langle\Fcal_{ci}\, \Fcal^{ci}\rangle =0
\end{equation}
for the Yang-Mills energy-momentum tensor $T^\ve_{\mu\nu}$ with $T^0_{ab}=\lim_{\ve\to 0} T^\ve_{ab}$. For the form of
Yang-Mills equations and the constraint equations $T^\ve_{ab}=0$ for $\ve >0$ see \cite{7}. In general, for $\ve\in [0,1]$ we
assume that fields $\Acal_\mu$ and $\Fcal_{\mu\nu}$ smoothly depend on $\ve$ and can be expanded in power series in $\ve$,
e.g. $\Acal_\mu=\Acal_\mu^0+\ve^2\Acal_\mu^1+\ve^4\Acal_\mu^2+...\ $. Note that $\Fcal^{\mu\nu}(\ve)$ should not be confused
with $\Fcal^{\mu\nu}_\ve = g_\ve^{\mu\sigma}g_\ve^{\nu\la}\Fcal_{\mu\nu}(\ve)$ in (\ref{8}). We omit $\ve$ from
$\Fcal_{\mu\nu}(\ve)$ for simplicity of notation. In (\ref{12})-(\ref{15}) we have zero terms in $\ve$ and omit index ``0"
from the fields. In fact, in (\ref{11}) we have $\Fcal_{ij}^0 =\partial_i\Acal_j^0 -\partial_j\Acal_i^0
 + [\Acal_i^0 , \Acal_j^0]$ but  $\Fcal_{ij}^1$, $\Fcal_{ij}^2$ etc. must not be zero.

\bigskip

\noindent {\bf 5. Flat connections. } Consider first the adiabatic  flatness equation (\ref{11}). Flat connection
$\Acal_{H^2}{:=}\Acal_i\diff x^i{=}\Acal_i(\ve{=}0)\diff x^i$ on $H^2$ has the form
\begin{equation}\label{16}
 \Acal_{H^2}=g^{-1}\hd g\with \hd =\diff x^i\pa_i\for\pa_i=\frac{\pa}{\pa x^i}\ ,
\end{equation}
where $g$ is a smooth map from $H^2$ into the gauge supergroup $G$ for any fixed $x^a\in\Si_2$. We impose on $g$ in (\ref{16})
the framing condition $g(x^3=1, x^4=0)=\Id$ (since constant $g$ in (\ref{16}) gives $\Acal_{H^2}\equiv 0$) and denote by
$C^\infty_0(H^2, G)$ the space of framed flat connections on $H^2$ given by (\ref{16}). On $H^2$, as on a manifold with
boundary, the (super)group of gauge transformations is defined as (see e.g. \cite{8, 6, 9})
\begin{equation}\label{17}
 \Gcal_{H^2}= \left\{g: H^2\to G\mid g^{}_{|\pa H^2}=\Id\right\}\ .
\end{equation}
Hence the solution space of the equation (\ref{11})  is the infinite-dimensional space $C_0^\infty (H^2, G)$ and the moduli
space  is the based loop (super)group (cf. \cite{6,8})
\begin{equation}\label{18}
 \Mcal = C^\infty_0 (H^2, G)/  \Gcal_{H^2}=\O G\ .
\end{equation}
This space can also be represented as $\O G = LG/G$, where $LG=C^\infty (S^1, G)$ is the loop supergroup with the circle
$S^1=\pa H^2$.

 \bigskip

\noindent {\bf 6. Moduli space. } On the moduli space $\Mcal=\O G$ we introduce coordinates $(X^\a_{(n)}, \th^{Ap}_{(n)})$ and
$(Y^\a_{(n)}, \chi^{Ap}_{(n)})$, where $\a , A, p$ run as before and $n\in\N$ appears from expanding coordinates in
$\sin(n\vp)$ and $\cos(n\vp)$ for $\vp\in S^1=\pa H^2$. We restrict ourselves to the subspace $G\subset \O G$ by putting
\begin{equation}\label{19}
(X^\a_{(1)}, \th^{Ap}_{(1)})= (X^\a , \th^{Ap}) \and
(Y^\a_{(1)}, \chi^{Ap}_{(1)})= -(X^\a , \th^{Ap})
\end{equation}
and assuming that all coordinates with $n\ne 1$ have zero values. Thus, our moduli space is  $G\subset \O G$.
In the adiabatic approach it is assumed that $\Acal_\mu = \Acal_\mu (x^a, x^i, X^\a , \th^{Ap})$ depend on
$x^a\in \Si_2$ only via moduli parameters \cite{10,11}, i.e. $\Acal_\mu = \Acal_\mu (X^\a (x^a), \th^{Ap}(x^a), x^i)$.
Then moduli of gauge fields define the map
\begin{equation}\label{20}
(X, \th ): \Si_2\to G \with  (X (x^a), \th (x^a))= \left\{X^\a (x^a), \th^{Ap}(x^a)\right\}\ ,
\end{equation}
where $G$ is now our moduli space . Acting by gauge transformations from (\ref{17}) on flat connections $\Acal_i$ in
(\ref{16}) which depend only on moduli $(X, \th )$ from (\ref{19}), we obtain the subspace $\Ncal$ in the full solution space
$C^\infty_0(H^2, G)$. The moduli space of these solutions is
\begin{equation}\label{21}
 G=\Ncal /\Gcal\ ,
\end{equation}
where $\Gcal = \Gcal^{}_{H^2}$ for any fixed $x^a\in \Si_2$.

The maps (\ref{20}) are constrained by the equations (\ref{13})-(\ref{15}). Since $\Acal^{}_{H^2}$ is a flat connection for any
$x^a\in \Si_2$, the derivatives  $\pa_a\Acal_i$ have to satisfy the linearized (around  $\Acal^{}_{H^2}$) flatness condition,
i.e.  $\pa_a\Acal_i$ belong to the tangent space $\Tcal_\Acal\Ncal$ of the space $\Ncal$. Using the projection $\pi : \Ncal\to G$ with fibres $\Gcal$,
one can decompose $\pa_a\Acal_i$ into the two parts
\begin{equation}\label{22}
T_\Acal\Ncal= \pi^*T_\Acal G\oplus T_\Acal \Gcal \quad\Leftrightarrow\quad\pa_a\Acal_i=\Pi_a^\a\xi_{\a i} +
(\pa_a\th^{Ap})\xi_{Api} + D_i\eps_a \ ,
\end{equation}
where
\begin{equation}\label{23}
\Pi_a^\a :=\pa_a X^\a - \im \de_{pq}\bar\th^p\ga^\a\th^q\ ,
\end{equation}
$\eps_a$ are $\gfrak$-valued gauge parameters ($D_i\eps_a\in T_\Acal \Gcal$) and  $\{\xi_{\a}=\xi_{\a i}\diff x^i ,
\xi_{Ap}=\xi_{Ap i}\diff x^i\}$ can be identified with $\gfrak =\,$Lie$\,G$. We will see in a moment that $(\xi_{\a i},
\xi_{Ap i})=(\xi_{\a}, \xi_{Ap})\xi_i$ with $(\xi_i)=(\sin\vp , -\cos\vp )$ mentioned in the Remark on p.1.

The gauge parameters $\eps_a$ are determined by the gauge fixing conditions
\begin{equation}\label{24}
  \de^{ij} D_i\xi^{}_{\Delta j}=0\quad\Rightarrow\quad \de^{ij} D_iD_j\eps_a=\de^{ij} D_i \pa_a\Acal_j \ ,
\end{equation}
where the index $\Delta$ means $\a$ or $Ap$. It is easy to see that
\begin{equation}\label{25}
  \de^{ij} D_i\xi_{\a j}=\de^{ij} \pa_i\xi_{\a j}=0\quad\Rightarrow\quad  \xi_{\a i}=\xi_\a\xi_i\with (\xi_i)=(\sin\vp , -\cos\vp )
\end{equation}
and similarly $\xi_{Ap i}=\xi_{Ap}\xi_i$. Another form of $\xi_i$ is $\xi_i=\ve_{ij}\pa_j r$ with $r^2=\de_{ij}x^i x^j$. It is easy to see that
$\de^{ij}\xi_i\xi_j=1.$

\bigskip

\noindent {\bf 7. Effective action. } Recall that $\Acal_i$ are given by (\ref{16}) and $\Acal_a$ are yet free. In the
adiabatic approach one choose $\Acal_a=\eps_a$ \cite{10, 11} and $\eps_a$ are defined from (\ref{24}). Then we obtain
\begin{equation}\label{26}
 \Fcal_{ai}=\pa_a\Acal_i - D_i\Acal_a = [\Pi_a^\b\xi_{\b} + (\pa_a\th^{Ap})\xi_{Ap}]\xi_i\in T_\Acal G\ .
\end{equation}
Substituting  (\ref{26}) into (\ref{13}), we see that  (\ref{13}) is resolved due to  (\ref{24}). Substituting  (\ref{26}) in
(\ref{14}), we will get the equations of motion for $X^\a (x^a), \th^{Ap}(x^a)$ which follow from the action (\ref{12}) which
after inserting  (\ref{26}) into  (\ref{12})  and integrating over $H^2$ becomes
\begin{equation}\label{27}
S_0=\int_{\Si_2} \diff x^1 \diff x^2\,\sqrt{|\det g^{}_{\Si_2}|}\, g^{ab}\,\Pi_a^\a\,\Pi_b^\b\,\eta_{\a\b}\ .
\end{equation}
This is the kinetic part of the Green-Schwarz superstring action. Note that
\begin{equation}\label{28}
 \eta_{\a\b}=\frac{1}{\pi}\int_{H^2}  \diff x^3 \diff x^4\,\langle\xi_\a\,\xi_\b\rangle \de^{ij}\xi_i\xi_j\ .
\end{equation}
As we mentioned in the item 3, this result does not depend on which metric ($g_{ij}=\de_{ij}$ or $g_{ij}$ from  (\ref{10})) we choose
on the disc $H^2$. Substituting  (\ref{26}) into the constraint equations  (\ref{15}) and integrating them over $H^2$, we obtain the equations
\begin{equation}\label{29}
 \eta_{\a\b} \,\Pi_a^\a\,\Pi_b^\b - \sfrac12\, g_{ab}\, g^{cd}\,\eta_{\a\b} \,\Pi_c^\a\,\Pi_d^\b =0\ ,
\end{equation}
which can also be derived from  (\ref{27}) by variation of the metric $g^{ab}\to\de g^{ab}$. Obviously, for $\theta^p=0$ one
gets the bosonic string action.

\bigskip

\noindent {\bf 8. Wess-Zumino-type term. } The action  (\ref{27}) is not yet the full Green-Schwarz action which contains
additional Wess-Zumino-type term \cite{12}. This term is described as follows. One considers a Lorentzian 3-manifold $\Si_3$
with the boundary $\pa\Si_3=\Si_2$ and coordinates $x^\ah, \ah =0,1,2$. On  $\Si_3$  one introduces the 3-form \cite{4}
\begin{equation}\label{30}
\O_3= \im\,\diff x^\ah\Pi_\ah^\a\wedge(\check\diff\bar\th^1\ga^\b\wedge\check\diff\th^1 -
\check\diff\bar\th^2\ga^\b\wedge\check\diff\th^2) \,\eta_{\a\b}=\check\diff\O_2\ ,
\end{equation}
where
\begin{equation}\label{31}
\O_2=-\im\check\diff X^\a\wedge(\bar\th^1\ga^\b\check\diff\th^1 - \bar\th^2\ga^\b\check\diff\th^2)\with \check\diff= \diff x^\ah\frac{\pa}{\pa x^\ah}\ .
\end{equation}
Then the term
\begin{equation}\label{32}
 S_{WZ} = \int_{\Si_3}\O_3 = \int_{\Si_2}\O_2
\end{equation}
is added to the functional (\ref{27}) and the Green-Schwarz action is
\begin{equation}\label{33}
  S_{GS}= S_{0} +S_{WZ}\ .
\end{equation}

To get the term (\ref{32}) from the Yang-Mills theory let us consider the 5-manifold $M^5=\Si_3\times H^2$ with coordinates
$x^\ah$ on $\Si_3$.
  Note that in addition to the components $\Fcal_{ai}$ in (\ref{26}) of Yang-Mills fields we now have the components
\begin{equation}\label{34}
\Fcal_{0i}=[(\pa_0X^\a - \im\,\de_{pq}\bar\th^p\ga^\a\pa_0\th^q)\xi_\a + (\pa_0\th^{Ap})\xi_{Ap}]\xi_i\ .
 \end{equation}
Notice that
\begin{equation}\label{35}
\Fcal_{\ah i}\xi^i=[(\pa_\ah X^\a - \im\,\de_{pq}\bar\th^p\ga^\a\,\th^q)\xi_\a + (\pa_\ah\th^{Ap})\xi_{Ap}]=:\omega_\ah
 \end{equation}
do not depend on $\vp$ since $(\xi_i)=(\sin\varphi, -\cos\varphi)$ is the unit vector on $H^2$ running over the boundary
$S^1=\pa H^2$ , $\xi_i \xi^i=1$. Let us introduce the one-forms $\omega =\omega_\ah\diff x^\ah$  and the Wess-Zumino-type
functional
\begin{equation}\label{36}
 S_{WZ} = \frac{1}{\pi}\int_{\Si_3\times H^2}f^{}_{\Gamma\Delta\Lambda}\, \omega^{\Gamma}\wedge
 \omega^{\Delta}\wedge\omega^{\Lambda}\wedge\diff x^3\wedge\diff x^4= \int_{\Si_3}\O_3
 = \int_{\Si_2}\O_2\ ,
\end{equation}
where $\O_3$ and $\O_2$ are the forms given by (\ref{30}),(\ref{31}) and the structure constants $f^{}_{\Gamma\Delta\Lambda}$
are written down in \cite{4}. Thus, adding the functional
\begin{equation}\label{37}
 \frac{1}{\pi}\int_{\Si_3\times H^2}\diff x^\ah\wedge \diff x^\bh\wedge \diff x^{\hat c}\wedge\diff x^3\wedge \diff x^4\,
 f^{}_{\Gamma\Delta\Lambda}\, \Fcal^{\Gamma}_{\ah i}\xi^i \Fcal^{\Delta}_{\bh j}\xi^j\Fcal^{\Lambda}_{\hat c k}\xi^k
\end{equation}
to the action (\ref{9}), we will get the Green-Schwarz superstring action in the adiabatic limit $\ve\to 0$.

\bigskip

\noindent {\bf 9. Concluding remarks.} We have introduced the Yang-Mills model on $\Si_2\times H^2$ whose action functional in
the low-energy limit reduces to the Green-Schwarz superstring action. Combining this result  with the result for bosonic
string \cite{7}, one can show that heterotic string theory can also be embedded into Yang-Mills theory as a subsector of
low-energy states. Thus, all five superstring  theories can be described in a unified manner via infrared limit of Yang-Mills
theory on $\Si_2\times H^2$. Such Yang-Mills models were not studied in the literature. But they may give a better base for
string unification than elusive M-theory and  $d{=}6$ superconformal field theory of multiple M5-branes.

The fibres of Yang-Mills bundle over $\Si_2\times H^2$ are the supermanifolds $G$ with Minkowski space $\R^{9,1}$ as the
bosonic part. Considering quantum Yang-Mills theory for small perturbations of Minkowski metric, one gets perturbative
description of quantum gravity which in the adiabatic limit will reduce to the stringy description. Put differently, embedding
superstring theory on $\Si_2$ into Yang-Mills theory on $\Si_2\times H^2$ allows one to raise the description of quantum
gravity from stringy level to the Yang-Mills level. Then the gravitation will be described via Yang-Mills fields for the
diffeomorphism structure group at least perturbatively.

Last but not least, quantum Yang-Mills theory is developed incomparably better than quantum string theory. Calculations in
quantum Yang-Mills theory are much easier than in string theory. Of course, to develop quantum Yang-Mills theory on
$\Si_2\times H^2$ with gauge supergroup $G$ is not a technically simple task. However, studying this Yang-Mills model can
improve understanding of many questions in string theories, e.g. those related with various dualities and AdS/CFT
correspondence.

\bigskip

\noindent {\bf Acknowledgements}

\noindent This work was partially supported by the Deutsche Forschungsgemeinschaft grant LE 838/13.

\end{document}